\documentclass[12pt]{iopart}
\usepackage{graphicx}
\begin{document}

\title[Multi-Step Ordering in Kagome and Square Artificial Spin Ice]{Multi-Step Ordering in Kagome and Square Artificial Spin Ice }   
\author{C.J. Olson Reichhardt$^{1}$,
A. Lib{\'a}l$^{2}$, and C. Reichhardt$^{1}$}
\address{
$^1$Theoretical Division and Center for Nonlinear Studies,
Los Alamos National Laboratory, Los Alamos, New Mexico 87545\\  
$^2$Faculty of Mathematics and Computer Science,
Babes-Bolyai University, RO-400591 Cluj-Napoca, Romania}
\ead{cjrx@lanl.gov}

\begin{abstract}
We show that in  colloidal models of artificial kagome and 
modified square ice systems, a variety of ordering and disordering regimes occur
as a function of biasing field, temperature, and colloid-colloid
interaction strength, including
ordered monopole crystals, biased ice rule states, 
thermally induced ice rule ground states, biased triple states,  and
disordered states. 
We describe the lattice geometries and 
biasing field protocols that create the 
different states 
and explain the formation of the states in terms of sublattice 
switching thresholds.
For a system prepared in a monopole lattice state,
we show that 
a sequence of different orderings occurs for increasing temperature.  
Our results also explain several features observed in nanomagnetic
artificial ice systems under an applied field. 
\end{abstract}
\pacs{82.70.Dd,75.10.Hk,75.10.Nr}
\maketitle

\section{Introduction}
Spin ices have been extensively studied 
as ideal systems that exhibit geometric frustration effects
since not 
all the pairwise spin interactions 
can be satisfied simultaneously \cite{Anderson,Gingras,Pauling,Anderson2}. 
Such systems are termed spin ices
due to their similarity with a water ice phase in which proton ordering into
corner-sharing tetrahedra is frustrated but obeys the ``ice rule'' of
two protons ``in'' (close to the O atom) and two protons ``out'' (far
from the O atom) in every tetrahedron
\cite{Pauling}. 
In spin ices, 
the corresponding spin ice rules determine how many spins point toward or
away from each vertex in the system, and these rules vary depending on
the geometry of the system.
For example, in two dimensional (2D)
square ice, ice-rule-obeying states have two spins pointing toward each
vertex and two spins pointing away, and it is possible for the spins to 
organize into an ordered
ground state configuration
\cite{Shiffer,Moller}. For 2D kagome ice, ice-rule-obeying states
have two spins in and one spin out out or two spins out and one spin in; 
in this case, the lowest energy state is not ordered but contains only vertices
that obey these ice rules.

There has recently been growing interest in 
2D artificial spin ice systems created using nanomagnetic
arrays 
\cite{Shiffer,Moller,Nisoli,Li,GB,EM,Zabel,OT,Ladak,Cumings,Mengotti,Tanaka,Mo,S,P}, 
2D colloidal particle assemblies \cite{Libal,Colloid}, or 
vortices in nanostructured superconductors \cite{Vortex}. 
In the nanomagnetic arrays, the orientation of the magnetic moment of each
nanoisland produces the effective spin ordering.
The colloidal and vortex systems more closely resemble the 
water ice system in that the
repulsive interactions of the colloids or vortices at vertices resemble 
the interactions between charged protons, and the ice rules indicate
the number of colloids or vortices that sit close to or away from a vertex.    
Experimental versions of these systems 
allow for direct imaging of the microscopic spin configurations as well as 
for careful control of numerous 
parameters that are not accessible in real spin systems.

In the work of Wang {\it et al.} \cite{Shiffer} 
on 2D square ice, 
as the interactions between neighboring nanomagnets increased, the
system was increasingly filled with ice-rule-obeying vertices;
however, the predicted ground state 
was not observed due to quenched disorder effects in the nanomagnet array.
Experiments on kagome artificial ice systems soon followed and showed
that the ice rules predicted for that geometry were obeyed \cite{Cumings}. 
In more recent square ice studies using more advanced techniques,
large regions of the sample adopted
the square ice 
ground state 
while non-ice-rule obeying vertices created grain boundaries \cite{GB}.
Such grain boundary formation 
was previously predicted in vortex simulations 
to occur for weak quenched disorder,
while for stronger disorder isolated
non-ice-rule defects begin to appear \cite{Vortex}. 

Lack of thermalization is 
an issue for the experimental nanomagnet systems 
since thermal activation of the spin configurations
is weak or nonexistent.
This can be partially
mitigated by applying a changing external field \cite{Nisoli,Li}. 
It was recently shown that 
biased states appear under applied external fields for both square and
kagome ice systems,
while cycled external fields can be used to generate
hysteresis curves \cite{OT,Zabel,EM,Ladak}. For kagome ice, 
Zabel {\it et al.} \cite{Zabel} 
observed an ordered monopole state where each vertex has either all spins
pointing out or all spins pointing in, with the vertices alternating sign
in a crystalline arrangement.
Other studies have examined the creation and motion
of monopole excitations in the presence
of a biasing field for both square and kagome ice systems \cite{OT,EM}.       
Recent theoretical studies have found that 
2D kagome ice can undergo a two-stage ordering
transition from a paramagnetic state to
an ice-rule-obeying state followed by an ordered monopole state 
\cite{Chern}. 
The thermal disordering or melting of the ordered states 
can be readily studied in colloidal artificial ices,
and recently materials have been 
created with a spin ordering temperature that is 
near room temperature, making it now possible to
melt magnetic artificial spin ices \cite{K}.

Here we show that a variety of different types of orderings are possible 
in both kagome and modified square ice systems
as a function of external field, temperature, and interaction strength
for a colloidal artificial spin ice system 
composed of arrays of elongated optical traps that each capture one colloid. 
Experimental studies have demonstrated frustration 
in buckled 2D colloidal layers \cite{Colloid} while
numerical studies have shown that charged colloids
in elongated trap arrays exhibit 
the same spin ice rules observed for nanomagnet arrays \cite{Libal}.  
Computational and experimental studies of colloidal ordering on 2D
square and triangular substrates have revealed numerous ordered states
and multi-step transitions between the states
\cite{Korda,Reichhardt,Frey,Trizac,Tizac,Brunner,Mangold,Mikael}. 
The elongated optical traps we consider here permit each colloid to
sit in one of two locations at either end of the trap.
Arrays
of this type of trap have been 
created experimentally to study various types of ratchet effects, and
preferential ordering of the colloids into a state that minimized the
colloid-colloid interactions was observed 
\cite{Babic}.
An advantage of the colloidal system over the nanomagnetic systems is that
thermal activation of the effective spin degrees of freedom is possible.
The relative importance of the thermal fluctuations can be varied either by
adjusting the temperature or by holding the temperature fixed and modifying
the strength of the colloid-colloid interactions.
It 
is also possible to 
bias the colloidal artificial ice system
using an electric field.    

\begin{figure}
\center
\includegraphics[width=3.5in]{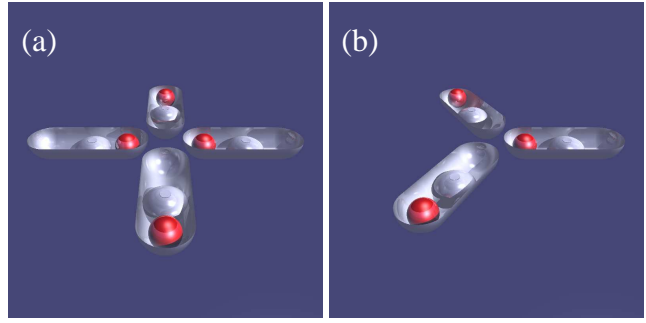}
\caption{
(a) Schematic of artificial square 
ice system consisting of elongated traps with two potential minima. 
An elementary unit or vertex 
consists of four traps that each capture one charged colloid. The effective
spin direction is defined to be toward the end of the trap
where the colloid is sitting.
Ice-rule obeying states have two colloids close to the vertex and 
two colloids away from the vertex; one of the two possible ground state
configurations is illustrated.
(b) Schematic of an artificial 
kagome ice system.  Ice-rule obeying states have either two colloids
close to the vertex and one away from the vertex or two colloids away
from the vertex and one close to the vertex, as shown.
}
\label{fig:1}
\end{figure}

{\it Simulation-}
We consider a 2D array of 
$N$ elongated traps that each contain a single colloid.
Each trap has two potential minima where the colloid can sit, as shown
in figure 1(a) where we illustrate a single vertex of the square
ice system.
The effective spin at each site
is defined to point toward the end of the trap occupied by the colloid.
The colloids interact with each other 
via a repulsive screened Coulomb or Yukawa potential
given by $V(r_{ij}) = r_{ij}^{-1}\exp(-\kappa r_{ij}){\bf {\hat r}}_{ij}$  
where $\kappa=4/a_0$ is the screening length and $a_{0}$ is 
the average spacing between the vertices. 
Since the colloids are repulsive it is
very energetically costly for all the colloids around a vertex to sit
close to the vertex in a positive monopole state;
on the other hand, the lowest energy state for a vertex is the negative
monopole configuration in which all the colloids sit away from the vertex.
Due to geometrical constraints
it is not possible for all of the vertices to adopt negative monopole
configurations.
The most cost-effective way to accommodate the 
colloid-colloid repulsion in the square ice
is for each vertex to 
have two close colloids and two far
colloids
in a ``two in-two out'' state that obeys the square ice rules.
These ice rule obeying states can be biased, with the close colloids
occupying traps oriented at 90$^\circ$ to each other, or they can be the
slightly lower energy ground state configuration shown in
figure~1(a) \cite{Libal}.
For kagome ice, shown schematically in figure 1(b),
the ice rule obeying vertices have either 
two in and one out or one out and two in.

The colloid dynamics evolve according to the following overdamped 
equation of motion: 
\begin{equation}
\eta \frac{d{\bf R}_{i}}{dt} = {\bf F}^{cc}_{i} + {\bf F}^{ext} + {\bf F}^{T}_{i} + 
{\bf F}^{s}_{i}   .
\end{equation}
Here $\eta$ is the damping constant and the colloid-colloid
interaction force is 
${\bf F}_{i}^{cc} = -F_0q^2\sum^{N}_{i\neq j}\nabla_i V(r_{ij})$ with
$F_{0} = Z^{*2}/(4\pi\epsilon\epsilon_{0})$,
$r_{ij}=|{\bf r}_{i} - {\bf r}_{j}|$,
and ${\bf {\hat r}}_{ij}=({\bf r}_{i}-{\bf r}_{j})/r_{ij}$.
${\bf r}_{i(j)}$ is the position of particle $i$($j$),
$Z^{*}$ is the unit of charge, 
$\epsilon$ is the dielectric constant of the solvent,
and $q$ is the magnitude of the charge on a single colloid. 
The strength of the repulsion between the colloids can be 
controlled by varying 
$q$.
The thermal force ${\bf F}^{T}$ arises from Langevin kicks with
$\langle {\bf F}^{T}_{i}\rangle = 0$ and 
$\langle {\bf F}_i(t){\bf F}_j(t^{\prime})\rangle = 2\eta k_{B}T\delta_{ij}\delta(t - t^{\prime})$.  
The substrate force ${\bf F}^{s}_{i}$ arises from $N$ traps composed
of two parabolic ends capping a cylindrical confining area of length
$l=1.333a_0$ and width $d_p=0.4a_0$
with a maximum strength of $F_{p}$ and radius $r_{p}$; an additional 
parabolic barrier of height $f_r$
is placed at the center of the trap to produce two 
potential minima at each end of the trap \cite{Libal}.  
The external biasing force is given by 
${\bf F}^{ext}=F_{ext}[\cos(\theta_{ext}){\bf \hat{x}} + \sin(\theta_{ext}){\bf \hat{y}}]$, with $\theta_{ext}=0$ for kagome ice and $\theta_{ext}=45^\circ$ for
square ice. 
For the nanomagnetic system, an in-plane applied external field was used
as a biasing force that could align the magnetic moments of the nanoislands.
For the colloidal system, the sample can be biased by an in-plane electric
field, while
for a system of vortices in a type-II superconductor the bias would come from
an in-plane applied current. 
In this work we consider external forces that are strong enough to induce
hopping over the central barrier of the traps but not strong enough
to cause the colloids to escape from the traps.
 
\begin{figure}
\center
\includegraphics[width=4.5in]{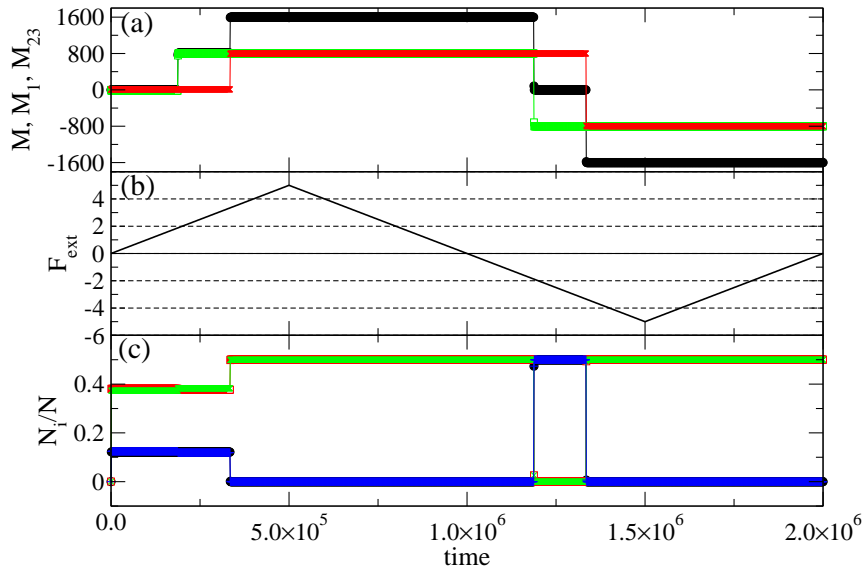}
\caption{ 
A kagome ice system with no colloid-colloid interactions ($q^2=0$) and
$f_r=4.0$
for $F_{ext}$ applied along
the $x$ axis.
(a) The effective magnetization $M$ for the entire sample (black filled
circles) vs time in simulation steps, along with the sublattice magnetizations
$M_1$ (green open squares) and $M_{23}$ (red x's).
(b) Simulation protocol:  
$F_{ext}$ vs time.
The system is initially driven into an
ice rule obeying state which transitions into a monopole
state when $F_{ext}$ becomes negative.     
(c) $N_i/N$, the fraction of vertices of type $N_i$,  vs time.
Black filled circles: $N_0/N$; 
red open squares: $N_1/N$; 
green x's: $N_2/N$;
blue +'s: $N_3/N$.
The monopole states are $N_{0}$ and $N_{3}$.  
}
\label{fig:diverge}
\end{figure}

\begin{figure}
\center
\includegraphics[width=3.5in]{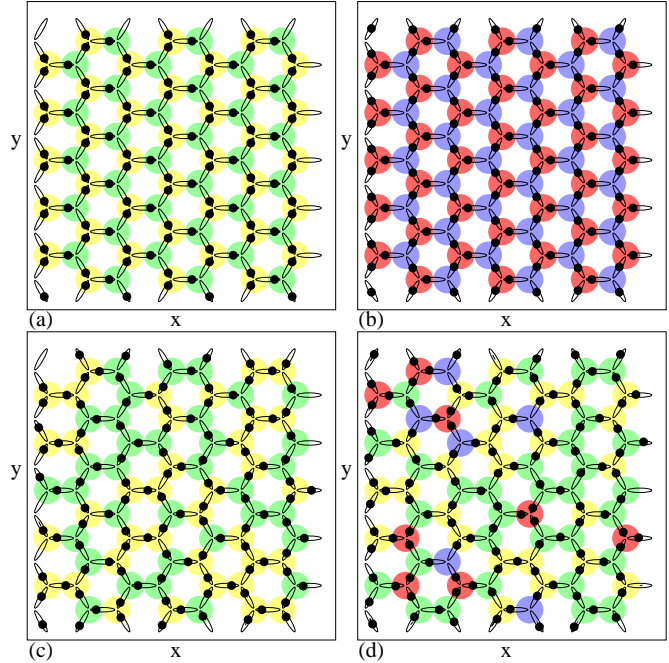}
\caption{
Black circles: particle locations; open ellipses: trap locations for
a $20a_0 \times 20a_0$ section of a kagome ice sample.  
Colored circles indicate vertex types:
$N_0$ (blue), $N_1$ (green), $N_2$ (yellow), and $N_3$ (red).
(a) The positive biased ice rule obeying state.
(b) The monopole state consisting of an ordered
lattice of $N_0$ and $N_3$ vertices.
(c) A finite temperature ice-rule obeying non-biased state. 
(d) A high temperature paramagnetic state where the vertex types are
uncorrelated.
}
\label{fig:vt}
\end{figure}

\section{Orderings for Kagome Ice}

We first consider a kagome ice array in an external field 
$F_{ext}$ applied along the $x$ axis, in alignment with 
the axes of 1/3 of the
traps in the sample.  In 
figure~2 
we plot the 
fraction of vertex types $N_i/N$, the effective magnetization $M$, and 
the applied external force $F_{ext}$ as a function of time
for a sample with noninteracting colloids at $q^2=0$.
The vertex types are defined as follows: 
$N_0$ has no colloids near the vertex; 
$N_1$ has one colloid near the vertex and two far from the vertex;
$N_2$ has two colloids near the vertex and one far;
and $N_3$ has three colloids near the vertex.  $N_1$ and $N_2$ vertices
obey the kagome ice rules, while $N_0$ and $N_3$ vertices are negative and
positive monopoles, respectively.

In figure 2(c), the noninteracting colloids begin in a randomized state
with $N_0/N = N_3/N = 0.12$ and $N_1/N = N_2/N = 0.38$.
As shown in figure 2(b), $F_{ext}$ is gradually increased from zero.
Just above $F_{ext} = 1.0$, the system jumps 
into the biased ice-rule obeying state illustrated in figure 3(a).
We define the contribution of each trap to the effective magnetization 
$M$ according to the component of the effective spin direction 
that is aligned with $F_{ext}$.
Each trap is assigned a value $s_i=\pm 1$ depending on whether it is biased
with or against the direction of $F_{ext}$, and this value is then weighted
by $\cos(\theta_i)$.  
The sublattice of trap sites that are aligned with the $x$ axis have
$\theta_i=0$ while the remaining trap sites 
that are oriented at $\pm 60^\circ$ 
to the $x$ axis 
have $\theta_i=60^\circ$.
In the biased state the net magnetization is 
$M/N_{\rm norm} = N_{\rm norm}^{-1}\sum_{i=1}^{N}s_i\cos(\theta_i)= 1.0$,
where $N_{\rm norm}=\cos(0^\circ)N/3 + \cos(60^\circ)2N/3$, and in figure 2(a)
the sample reaches the biased state at $F_{ext}=3.7$.  

\begin{figure}
\center
\includegraphics[width=4.5in]{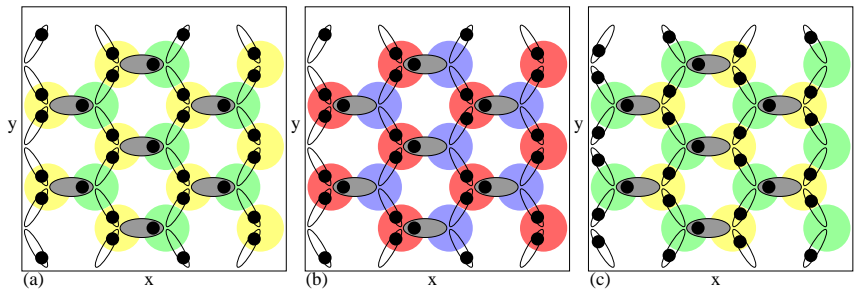}
\caption{
Schematic of the formation of the ordered monopole state in a kagome ice
sample under a biasing drive applied along the $x$ axis.  Black circles
are colloid positions and ellipses indicate trap locations. 
The traps can be broken into two sublattices: sublattice $A$ aligned with
the $x$ axis (shaded ellipses) and sublattice $B$ oriented at $\pm 60^{\circ}$
from the $x$ axis (open ellipses).
For noninteracting colloids ($q^2=0$) the switching or coercive field for
sublattice $A$ is lower than for sublattice $B$.
Colored circles indicate vertex types:
$N_0$ (blue), $N_1$ (green), $N_2$ (yellow), and $N_3$ (red).
(a) The system begins in the positively biased ice rule obeying state.
(b) As $F_{ext}$ becomes increasingly negative, it crosses the coercive
field of sublattice $A$ first.  Colloids in sublattice $A$ switch while those
in sublattice $B$ do not, creating the ordered monopole state.
(c) As $F_{ext}$ continues to decrease, sublattice $B$ switches and the sample
reaches the negatively biased ice rule obeying state.
}
\label{fig:vtb}
\end{figure}

After the
sample is fully biased, we decrease $F_{ext}$ as shown in figure 2(b).
At $F_{ext}=-1.7$, the biased state
switches into an ordered monopole state 
where $N_0$ and $N_3$ each jump from $0\%$ of the population to $50\%$ of the 
population while $N_1$ and $N_2$ simultaneously drop to zero
as shown in figure 2(c).
The ordered monopole state, illustrated in figure 3(b), 
is a lattice of positive and negative monopoles 
and is the same state observed in experiments 
on kagome nanomagnetic systems under external drives \cite{Zabel}. 
The ordered monopole state arises due to the fact that the kagome
lattice is effectively composed of 
two sublattices that have different switching
fields under an external $x$-direction drive.
This is illustrated schematically in figure 4 where the darker ellipses
indicate the sublattice with a lower switching field, which we term sublattice
$A$.
For the case of non-interacting colloids with $q^2=0$, 
the switching field for sublattice $A$ is simply the maximum 
force $f_r$ of the barrier in the center of the trap. Since the applied drive
is not aligned in the direction of the traps of 
sublattice $B$, the switching force for sublattice $B$ has a higher 
value of $f_{r}/\cos(60^{\circ})$. 
When we apply a sufficiently large
$F_{ext}$ in the positive $x$-direction to cause both the $A$ and $B$ 
sublattices to switch, the system enters the positively biased state illustrated
in figure 4(a).
After we reduce $F_{ext}$ back to zero, we begin applying $F_{ext}$ in the
negative $x$ direction.
Sublattice $A$ switches in the negative $x$ direction first due to its
lower threshold, but since the colloids in sublattice $B$ have not yet
switched,
the ordered monopole state forms as shown in figure 4(b).
For $q^2 \ne 0$ or finite colloid-colloid interactions, creation of
monopoles becomes energetically unfavorable, so for sufficiently high
$q^2$ the switching of sublattice $A$ simultaneously induces a switching
of sublattice $B$, bringing the system directly into the negatively
biased state illustrated in figure 4(c) and
preventing the formation of the ordered monopole
state.
For a certain range of nonzero $q^2$,
however, 
the ordered monopole state can still be stabilized 
when $F_{ext}$ passes through zero.
Our results suggest that in the experiments 
of Zabel {\it et al.} \cite{Zabel} the magnetic islands are in the 
weakly interacting or noninteracting limit. 

\begin{figure}
\center
\includegraphics[width=3.5in]{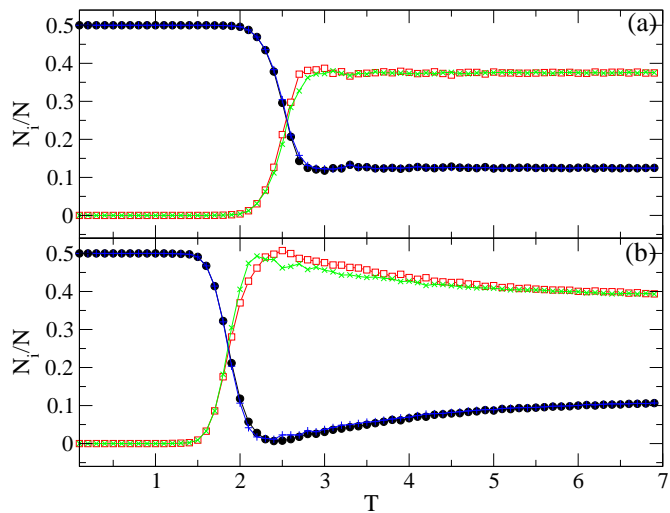}
\caption{
$N_i/N$ vs temperature $T$ in kagome ice samples with $f_r=1$
that have been initialized in the ordered monopole state; $F_{ext}$ is set
to zero before the temperature is applied.
Black filled circles: $N_0$; red open squares: $N_1$; green x's: $N_2$;
blue $+$'s: $N_3$.
(a) A sample with zero pairwise interactions $q^2=0$. 
Here the ordered monopole
state melts directly into a paramagnetic state. 
(b) A sample with $q^2=0.3$.
As $T$ increases,
the monopole density drops to zero while
$N_1/N$ and $N_2/N$ pass through peak values of 0.5,
indicating that the ordered monopole state
melts into an ice-rule obeying state. 
At higher temperatures, monopoles reappear while
$N_{1}/N$ and $N_{2}/N$ decrease and
approach the
value expected for the paramagnetic state.
}
\label{fig:vta}
\end{figure}

We next examine the thermal disordering of the 
ordered monopole state for varied colloid-colloid interaction strengths.       
In figure~5(a) we plot the vertex populations $N_{i}/N$ versus 
temperature $T$ for 
a noninteracting sample with $q^2=0$, while
figure~5(b) shows $N_i/N$ versus $T$
for an interacting sample with $q^2=0.3$.
In each case we set $F_{ext}=0$ before applying the temperature.
In the noninteracting system in figure 5(a), the ordered monopole state 
persists up to 
$T = 2.5$, 
at which point the $N_{i}/N$ cross over to the values 
expected for an uncorrelated random arrangement or paramagnetic
state. 
In figure 5(b) for $q^2 = 0.3$, 
$N_{3}/N$ and $N_{0}/N$ drop to zero with increasing $T$ while
$N_{1}/N$ and $N_{2}/N$ increase to 
$0.5$, indicating a crossover from 
the ordered monopole state into an ice-rule obeying state. 
This is not the biased ordered state illustrated in figure 3(a); instead,
the ice-rule obeying state lacks true long range order, as shown
in figure 3(c).
As $T$ is further increased, $N_{3}/N$ and $N_{0}/N$ gradually increase while
$N_{1}/N$ and $N_{2}/N$ decrease to values
close to those expected for a random or paramagnetic state of the type 
illustrated in figure 3(d).

\begin{figure}
\center
\includegraphics[width=4.5in]{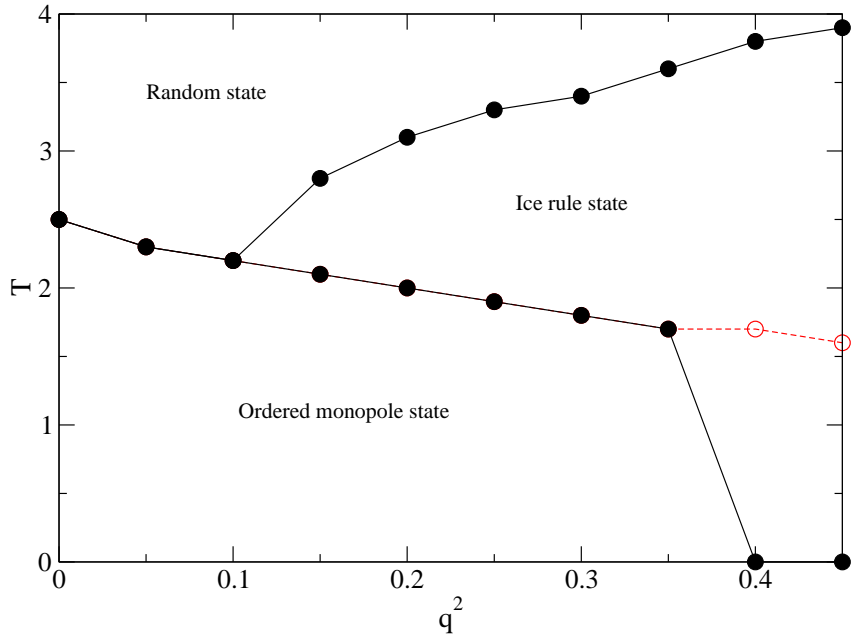}
\caption{ 
Phase diagram for $T$ vs pairwise interaction strength
$q^2$ for the kagome ice system 
with $f_r=1$ indicating
regions where the ordered monopole, ice-rule obeying, and
random or paramagnetic states appear.
For weak interactions (small $q^2$) the
monopole state disorders directly into the random state, while 
the width of the ice-rule obeying region grows with increasing $q^2$.
Red symbols: beyond the highest value of $q^2$ where the ordered monopole
state can be created with a biased drive protocol, 
the system can be artificially placed into the
ordered monopole state and then melted into the ice-rule obeying state.
}
\label{fig:diverge2}
\end{figure}

By conducting a series of simulations, we map out the 
different orderings as a function
of temperature and pairwise interactions as shown in 
figure~6. The line separating the ice rule obeying state
from the random state is defined as the point at which
$N_{1}/N$ and $N_{2}/N$ reach values that are within 10\% of those
expected for a random vertex arrangement.
For weak interactions or small $q^2$ the ordered monopole state
melts directly into the
random state. As $q^2$ increases, 
the ice-rule obeying region increases in extent 
while the line marking the crossover
between the ordered monopole state and the ice-rule obeying state drops
to lower $T$.
These results suggest that 
the ordered monopole states can only be observed 
for systems with effective spin-spin interactions that are weak compared
to the strength of the switching or coercive field.
For $q^2 > 0.4$ the ordered monopole state can no longer be prepared 
by means of the external field protocol since the repulsion between colloids 
becomes strong enough that switching of the $A$ sublattice immediately
induces switching of the $B$ sublattice.
Nevertheless, the ordered monopole state is still stable
at lower $T$ for
$q^2>0.4$ as can be seen by artificially preparing the lattice in the
ordered monopole state and then increasing the temperature.  
Such artificial preparation could be achived
in colloidal systems by, for example, tuning
the colloid-colloid interactions 
to a lower value in order to create the ordered monopole state
and then increasing the interactions to a higher value; 
however, it is generally not possible
to vary the interaction strength in magnetic systems.
The line
marking the melting into the ice-rule obeying state obtained by using
artificially created ordered monopole states is marked with a dashed
line in figure 6.

\begin{figure}
\center
\includegraphics[width=4.5in]{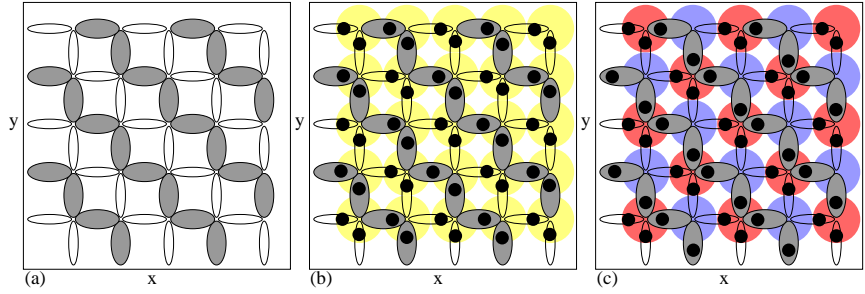}
\caption{ 
(a) Schematic of a modified square ice sample containing two trap sublattices
$A$ (shaded ellipses) with central barrier $f_r^1$ and $B$ (open ellipses)
with central barrier $f_r^2$.  We take $f_r^2/f_r^1=2$ so that sublattice $A$
has a lower threshold for switching.
$F_{ext}$ is applied along a line oriented at $45^\circ$ to the $x$ axis.
(b) A positively biased ice rule obeying state is formed by initial
driving in the positive direction, $\theta_{ext}=45^\circ$.  Yellow circles:
positively biased $N_2^b$ vertices.
(c) When the driving direction is switched to 
the negative direction, 
sublattice $A$ flips first producing a checkerboard monopole state. 
Blue circles: $N_0$ vertices; red circles: $N_4$ vertices.
The colloid-colloid interactions must be weak in order for the 
ordered monopole state to remain stable.
}
\label{fig:diverge3}
\end{figure}

\begin{figure}
\center
\includegraphics[width=4.5in]{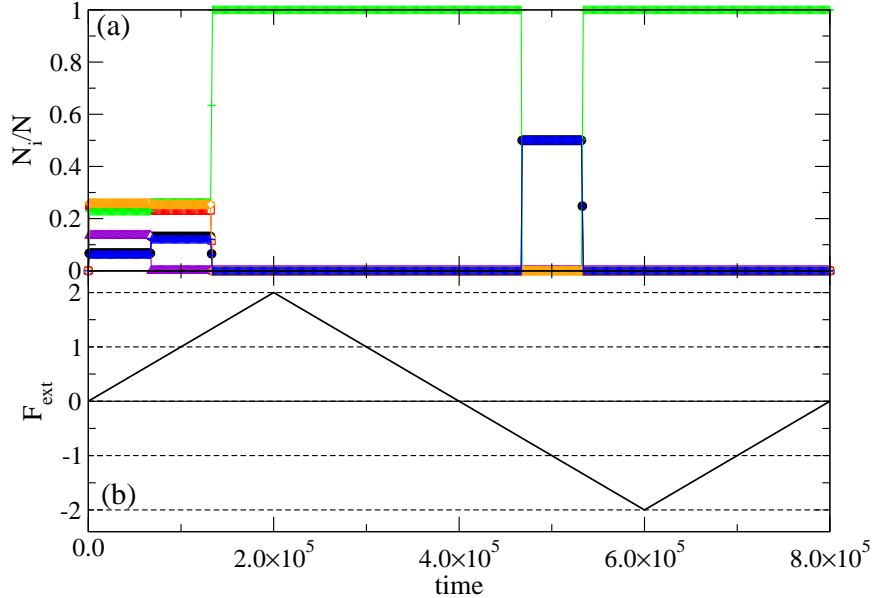}
\caption{ 
A modified square ice system as illustrated in figure 7
with $q^2=0.4$, $f_r^1=1.0$, and $f_r^2=2.0$ for $F_{ext}$ applied along
a line tilted at $45^\circ$ to the $x$ axis.
(a) $N_i/N$, the fraction of vertices of type $N_i$, vs time in simulation
time steps.  Black filled circles: $N_0$; red open squares: $N_1$; green 
$+$'s: biased $N_2^{b}$ vertices; purple open triangles: ground state $N_2^{gs}$
vertices; orange open diamonds: $N_3$; blue $x$'s: $N_4$.  The monopole
states are $N_0$ and $N_4$.
(b) $F_{ext}$ vs time.
The system switches into a positively biased ice rule obeying state 
with $N_2^{b}/N=1.0$ near
$F_{ext}=1.3$.  After the drive has been reversed, the system switches into
an ordered monopole state with $N_0/N=N_4/N=0.5$ at $F_{ext}=-0.65$ and then
orders into a negatively biased ice rule obeying state
at $F_{ext}=-1.3$.
}
\label{fig:divergeO}
\end{figure}

\section{Orderings For Square Ice}
In order to create an easily 
accessible ordered monopole state for a square ice lattice, we propose a
simple extension of the standard square ice geometry 
where we now have two sublattices as illustrated
in figure 7. 
Each sublattice has a different strength of the central barrier, $f_r^1$ and
$f_r^2$, and we take $f_r^2/f_r^1=2$.
In figure 7(a), the darker traps have the weaker barriers. 
Such a geometry could be produced in the nanomagnetic system by using two
different sizes or two different materials for the nanoislands, producing
two different coercive field values.
We apply a biasing force along a line oriented at $45^{\circ}$ to the $x$
axis.
In the square ice array, the vertex types $N_0$, $N_1$, $N_2$, $N_3$, and $N_4$
have 0, 1, 2, 3, and 4 colloids close to the vertex, respectively.
The $N_2$ vertices are further subdivided into biased vertices $N_2^b$ where
the two close colloids are in adjacent traps
as in figure 7(b), and ground state vertices
$N_2^{gs}$ where the two close colloids are on opposite sides of the vertex
as in figure 1(a).
In figure~8 we plot the the external drive $F_{ext}$ and fraction of vertex types
$N_j/N$ versus time. The system begins in a random state
with $N_{4}/N + N_{0}/N = N_2^{gs}/N = 0.125$ 
and $N_{1}/N = N_{2}/N = N_{3}/N = 0.25$.  
As $F_{ext}$ increases past $F_{ext}=1.3$, 
the sample enters the positively biased ice rule obeying state 
illustrated in figure 7(b), with $N_2^b/N=1$. 
When the external drive reaches $F_{ext}=2.0$, we begin decreasing $F_{ext}$
all the way to $F_{ext}=-2.0$.
The biased state persists until
$F_{ext} = -0.67$, at
which point 
a checkerboard monopole state appears as illustrated in figure~7(c).
In analogy to the kagome ice system, in the two-sublattice square ice
sample the monopole state forms when sublattice $A$, with its weaker
barrier, switches before sublattice $B$.
If the colloid-colloid interaction strength
is small enough, the particles in sublattice 
$B$ remain in their original positions, producing
the monopole state shown in figure~7(c). 
As $F_{ext}$ is further increased in the negative direction, sublattice $B$
switches 
at $F_{ext}=-1.3$ 
to form a negatively biased ice rule obeying state aligned with
the drive.
We note that in our previous work on square ice with only a single trap
sublattice, monopole formation was rare due to its high
energetic cost.

\begin{figure}
\center
\includegraphics[width=4.5in]{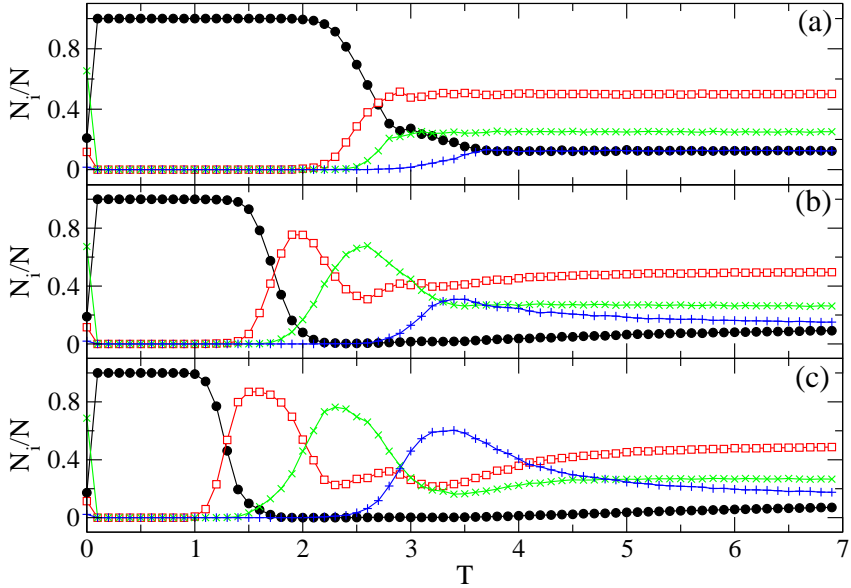}
\caption{ 
The vertex populations $N_i/N$ vs temperature $T$ in a modified square
ice system with $f_r^1=1.0$ and $f_r^2=2.0$ that has been prepared in a 
monopole state with $F_{ext}$ set to zero.
Black filled circles: monopole states $N_0/N+N_4/N$; red open squares:
$N_1/N + N_3/N$; green $x$'s: $N_2^{b}/N$ biased vertices;
blue $+$'s: $N_2^{gs}/N$ ground state vertices.
(a) A sample of noninteracting colloids with $q^2=0$ passes directly from
a monopole state to a disordered state.
(b) A sample with $q^2 = 0.2$ shows the formation of the
biased triple state followed by the thermally induced biased 
ice rule obeying state and then a random or paramagnetic state.
(d) In a sample with $q^2 = 0.4$, the monopole state 
is followed by the biased triple state, the 
thermally induced biased ice rule obeying state,
the thermally induced ground state, and the paramagnetic state.  
}
\label{fig:divergen}
\end{figure}

\begin{figure}
\center
\includegraphics[width=5.5in]{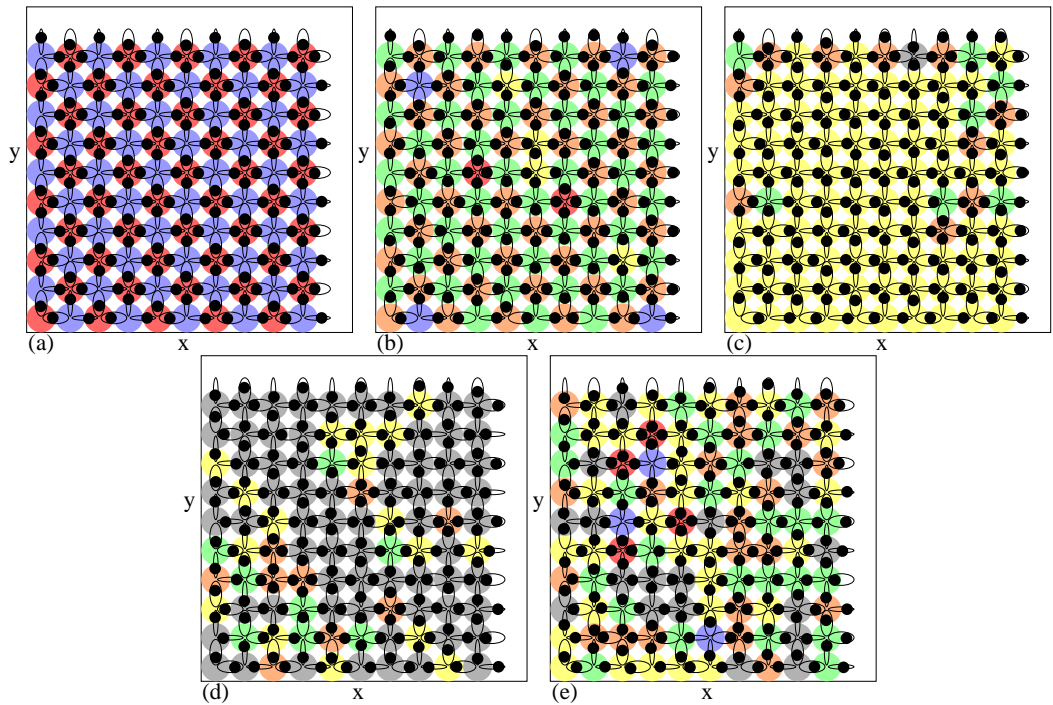}
\caption{ 
Black circles: particle locations; open ellipses: trap locations for a
$20a_0 \times 20a_0$ section of a modified square ice system
with $f_r^1=1.0$, $f_r^2=2.0$, and $q^2=0.4$.  The two
trap sublattices are indicated by ellipses of different sizes.  Colored
circles indicate vertex types: $N_0$ (blue), $N_1$ (green), 
biased $N_2^{b}$ (yellow),
ground state $N_2^{gs}$ (gray), $N_3$ (orange), and $N_4$ (red).
(a) The ordered monopole state. (b) The biased triple state.
(c) The thermally induced biased ice rule obeying state. 
(e) The thermally induced ground state. (f) The
paramagnetic or uncorrelated state.  
}
\label{fig:divergem}
\end{figure}

We next consider thermal effects on the ordered monopole 
state.  To prepare the sample,  
we sweep $F_{ext}$ to a value at which the monopole state appears and then
switch off $F_{ext}$.  If the pairwise interactions between colloids are not
too strong, the monopole state remains stable even without a biasing drive.
Figure~9(a) shows the vertex populations 
$N_i/N$ versus temperature $T$ for a system of noninteracting
colloids with $q^2=0$.
Here the monopole state 
illustrated in figure 10(a) disorders into the paramagnetic state
illustrated in figure 10(e), with
$N_0/N$ and $N_4/N$ monotonically dropping from $0.5$ to around $0.06$ but
never falling below this value.
In samples with interacting colloids, such as the system with $q^2=0.2$ shown
in figure 9(b),
$N_0/N$ and $N_4/N$ both drop from 0.5 to 0 
with increasing $T$ indicating that all monopoles
disappear from the system.  As $T$ is increased further, $N_0/N$ and $N_4/N$
gradually increase back to the paramagnetic value of 0.0625.
For $1.9 \leq T \leq 2.3$, where $N_0/N$ and $N_4/N$ are dropping 
to zero,
the values of $N_1/N$ and $N_3/N$ pass through a peak.  This is followed
by a window $2.3 \leq T \leq 3.3$ where $N_1/N$ and $N_3/N$ drop again 
while the value of $N_2^{b}/N$  peaks.
A weak peak in $N_2^{gs}/N$ appears in the vicinity of $T \sim 3.5$, while at
higher temperatures the vertex populations approach the values expected
in a completely random configuration.

The same trends are more clearly evident in figure 9(c) for a sample with
larger pairwise interactions, $q^2=0.4$.
Here the highest value reached by $N_1/N + N_3/N$ is nearly 0.9, indicating
that large portions of the sample are filled with $N_1$ or $N_3$ vertices.
The corresponding colloid configuration is illustrated in figure 10(b),
where local ordering of the $N_1$ and $N_3$ vertices into a lattice can
be seen.
We term this state a biased triple state; it forms due to the preferential
switching of colloids in the weaker sublattice $A$ traps.
As $T$ increases, $N_1/N$ and $N_3/N$ drop while $N_2^{b}/N$ peaks at
a value of $0.8$ corresponding to 
the thermally induced biased ice rule obeying state,
illustrated in figure 10(c).
At higher $T$, $N_2^b/N$ decreases again while $N_2^{gs}/N$ reaches its
peak value of 0.7 corresponding to the thermally induced ground state shown in
figure 10(d).
For still higher $T$, the $N_i/N$ approach the values expected for a random
configuration and the sample enters a paramagnetic state, illustrated
in figure 10(e).

The sequence of ordered states we observe for finite pairwise interactions
under finite temperatures can be understood by considering the interaction
of the metastable monopole state with the two trap sublattices.
Our protocol of sweeping $F_{ext}$ into
the monopole regime and then switching off $F_{ext}$ at zero temperature
causes the sample to be trapped in this metastable state for low
temperatures.
As $T$ increases from zero, 
the system first lowers its energy by moving one colloid out of each $N_4$
vertex into an $N_0$ vertex, creating $N_1$ and $N_3$ vertex pairs and
generating the biased triple state.
This produces the peak in $N_1/N$ and $N_3/N$ near $T=1.6$ in figure 9(c),
while figure 10(b) indicates that due to the finite temperature,
some non-$N_1$ or $N_3$ vertices still exist in the sample.
In the biased triple state,
the lower energetic cost of an $N_3$ vertex compared to an $N_4$ vertex 
means that a higher thermal activation energy is required to destroy 
$N_3$ vertices compared to $N_4$ vertices.  The difference in thermal 
activation energies determines the size of the temperature window where the
biased triple state remains metastable.
Once the temperature is large enough, colloids jump out of $N_3$ vertices
into neighboring $N_1$ vertices, resulting in the formation of $N_2$ vertices.
This hopping is more likely to occur in the weaker traps of sublattice
$A$, resulting in the preferential formation of biased $N_2^b$ vertices.
As a result, $N_2^b/N$ peaks near $T=2.3$ in figure 9(c) and the sample
enters the biased ice rule obeying configuration (with thermal defects) shown
in figure 10(c).
Since the $N_2^b$ vertices have a slightly higher energy than the ground state
$N_2^{gs}$ vertices \cite{Libal},
at even higher temperatures 
additional hopping produces an increased fraction of
$N_2^{gs}$ vertices, placing the system in the thermally induced and
thermally defected square ice ground state configuration illustrated
in figure 10(d).
For sufficiently high $T$, the correlations between neighboring traps are
lost and the sample enters the paramagnetic state shown in figure 10(e).

\begin{figure}
\center
\includegraphics[width=4.5in]{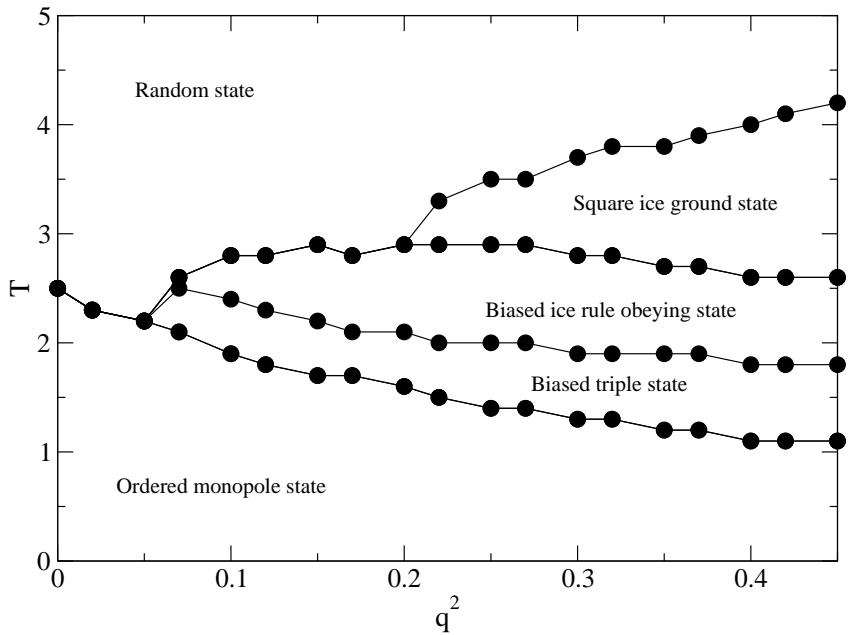}
\caption{ 
Phase diagram for $T$ vs $q^2$ for the modified square ice sample with 
$f_r^1=1.0$ and $f_r^2=2.0$ indicating regions where the ordered monopole,
biased triple, biased ice rule obeying, square ice ground state, and random
states appear.
As $q^2$ increases, the width of the ordered monopole state decreases while
the transition to the random state shifts to higher $T$.
}
\label{fig:divergeo}
\end{figure}

In figure~11 we plot a phase diagram of the regions in $T$ and $q^2$ space
where the different ordered states appear.
The lines indicate crossovers and are determined based on when the
different $N_i/N$ cross threshold values.
As $q^2$ increases,
the destruction of the ordered monopole state drops to lower values of $T$.
The biased triple state and biased ice rule obeying state maintain roughly
constant widths in $T$ as $q^2$ increases, although both phases shift to
lower ranges of $T$.
In contrast, the square ice ground state, which appears only for larger
values of $q^2$, increases in width as $q^2$ increases.
Above $q^2=0.45$ we can no longer stabilize the ordered monopole state
using our biasing protocol.
Our results indicate that
a simple modification of the square ice geometry 
can produce a wealth of orderings
in the artificial spin ice system.                 

We note that the phase diagrams for the kagome and 
modified square ice presented here do not exhibit true phases since we have
driven the system into a metastable state. 
In many of the nanomagnetic systems that exhibit
hysteresis, most of the 
observed states are also metastable unless a successful annealing
protocol has been applied.
All of our samples contained no quenched disorder.  
We expect that many of the states we observe would be gradually washed
out if increasing amounts of quenched disorder are added; 
however, there should still be observable correlations in the different states
that can be detected by analyzing the vertex population densities.

\section{Summary}
We have described the creation of different 
types of orderings in artificial kagome and modified square ice systems
using external fields, temperature, and particle-particle
interaction strength. 
We show that multiple-step ordering-disordering transitions 
occur that can be identified by counting the vertex populations.
In the kagome ice,
an ordered monopole state can be induced by applying a biasing field to
samples with pairwise interactions that are not too strong.  With
increasing temperature, the system crosses into an ice rule obeying state
and then into a high temperature paramagnetic state.
For stronger pairwise interactions, the monopole state is unstable
to the formation of the ice rule obeying state. 
Our results 
suggest that recent experimental observations of monopole ordering
in kagome nanomagnetic systems 
were performed in a weakly interacting or noninteracting regime.   
For square ice, we make a simple modification to the geometry by 
introducing two sublattices of traps that have different switching fields
and show that under the application of an external biasing field, this system
can form a checkerboard ordered monopole state.
The modified square ice exhibits multi-state ordering as a function of 
temperature starting from the monopole state, passing through
a biased triple state, a thermally induced biased ice rule obeying state, 
a thermally induced square ice ground state, and a disordered or
paramagnetic state.
Although our results are obtained specifically 
for colloids on periodic trap arrays, the behavior should be generic
to other artificial ice systems under 
external drives, including nanomagnetic array systems.  

\section{Acknowledgements}
This work was carried out under the auspices of the 
NNSA of the 
U.S. Department of Energy
at 
Los Alamos National Laboratory
under contract number
DE-AC52-06NA25396.
The work of A. Lib{\' a}l was supported by a grant of the Romanian National
Authority for Scientific Research, CNCS--UEFISCDI, project number
PN-II-RU-TE-2011-3-0114.

\end{document}